\documentclass[journal=jacsat,manuscript=article,layout=twocolumn]{achemso}

\usepackage{graphicx}% Include figure files
\usepackage{dcolumn}% Align table columns on decimal point
\usepackage{bm}% bold math
\usepackage{hyperref}% add hypertext capabilities
\usepackage[mathlines]{lineno}% Enable numbering of text and display math
\usepackage{amsmath}
\usepackage{caption}
\usepackage{subcaption}
\usepackage{color}
\usepackage{import}
\usepackage{xcolor}
\usepackage{soul}

\newcommand{\tauc}{\tau_{\text{c}}}
\newcommand{\onlinecite}[1]{\hspace{-1 ex} \nocite{#1}\citenum{#1}}
\let\oldmaketitle\maketitle
\let\maketitle\relax

\title{Beyond 1 ms Charge-Carrier Recombination Dynamics in the CsPbBr$_3$ Perovskite}

\author{Andr\'{a}s~Bojtor}
\affiliation{{Department of Physics, Institute of Physics, Budapest University of Technology and Economics, Műegyetem rkp. 3., H-1111 Budapest, Hungary}}
\alsoaffiliation{{Semilab Co. Ltd., Prielle Kornélia u. 4/a, 1117 Budapest, Hungary}}

\author{D\'{a}vid~Kriszti\'{a}n}
\affiliation{{Department of Physics, Institute of Physics, Budapest University of Technology and Economics, Műegyetem rkp. 3., H-1111 Budapest, Hungary}}
\alsoaffiliation{{Semilab Co. Ltd., Prielle Kornélia u. 4/a, 1117 Budapest, Hungary}}

\author{Ferenc~Kors\'{o}s}
\affiliation{{Semilab Co. Ltd., Prielle Kornélia u. 4/a, 1117 Budapest, Hungary}}

\author{S\'{a}ndor~Kollarics}
\affiliation{{Institute for Solid State Physics and Optics, Wigner Research Centre for Physics, PO. Box 49, H-1525, Hungary}}
\alsoaffiliation{{Department of Physics, Institute of Physics and ELKH-BME Condensed Matter Research Group Budapest University of Technology and Economics, M\H{u}egyetem rkp. 3., H-1111 Budapest, Hungary}}

\author{Thomas~Pinel}
\affiliation{{Department of Physics, Institute of Physics, Budapest University of Technology and Economics, Műegyetem rkp. 3., H-1111 Budapest, Hungary}}
\alsoaffiliation{{\'{E}cole Nationale Sup\'{e}rieure d'ing\'{e}nieurs de Caen, 6 Bd Mar\'{e}chal Juin, 14000 Caen, France}}

\author{M\'{a}rton~Koll\'{a}r}
\affiliation{{KEP Innovation Center, Ch. du Pré-Fleuri 5, 1228 Plan-les-Ouates, Switzerland}}

\author{Endre~Horv\'{a}th}
\affiliation{{KEP Innovation Center, Ch. du Pré-Fleuri 5, 1228 Plan-les-Ouates, Switzerland}}

\author{Xavier~Mettan}
\affiliation{{KEP Innovation Center, Ch. du Pré-Fleuri 5, 1228 Plan-les-Ouates, Switzerland}}

\author{Bence~G.~M\'{a}rkus}
\affiliation{{Stavropoulos Center for Complex Quantum Matter, Department of Physics and Astronomy, University of Notre Dame, Notre Dame, Indiana 46556, USA}}
\alsoaffiliation{{Institute for Solid State Physics and Optics, Wigner Research Centre for Physics, PO. Box 49, H-1525, Hungary}}
\alsoaffiliation{{Department of Physics, Institute of Physics and ELKH-BME Condensed Matter Research Group Budapest University of Technology and Economics, M\H{u}egyetem rkp. 3., H-1111 Budapest, Hungary}}

\author{L\'{a}szl\'{o}~Forr\'{o}}
\affiliation{{Stavropoulos Center for Complex Quantum Matter, Department of Physics and Astronomy, University of Notre Dame, Notre Dame, Indiana 46556, USA}}

\author{Ferenc~Simon}
\affiliation{{ELKH-BME Condensed Matter Research Group, Budapest University of Technology and Economics, Műegyetem rkp. 3., H-1111 Budapest, Hungary}}
\alsoaffiliation{{Institute for Solid State Physics and Optics, Wigner Research Centre for Physics, PO. Box 49, H-1525, Hungary}}

\email{simon.ferenc@ttk.bme.hu}

\begin{document}

\twocolumn[
\begin{@twocolumnfalse}
\oldmaketitle
\begin{abstract}
Knowledge of the charge-carrier recombination lifetime, $\tauc$, is crucial for the various applications of photovoltaic perovskites. We studied the novel inorganic perovskite, CsPbBr$_3$ and we observe recombination dynamics beyond $1$ ms below $200$ K and $\tauc$ approaching $100$ $\mu$s at room temperature. Time-resolved microwave-detected photoconductivity decay (TRMCD), used in combination with injection dependence, evidence that $\tauc$ is dominated by impurity charge trapping. The observed injection dependence is well corroborated by modeling of the trap mechanism. The ultra-long decay time is also consistent with photoconductivity measurements with a continuous-wave excitation at powers corresponding to around one Sun irradiation. While in principle charge-carrier trapping may limit the photovoltaic efficiency in single-cell photovoltaic devices, it could also lead to enhanced efficiency in tandem cells as well as for alternative applications including photodetection and quantum information storage. 
\end{abstract}
\end{@twocolumnfalse}]

%\maketitle


\twocolumn

\section*{Introduction}

Metal-halide perovskites have become the focus of research in a wide range of applications due to their low fabrication cost and the relative ease of their growth as compared to silicon while possessing extraordinary photonics properties \cite{ScienceReview,efficiency,PerovskiteLED_NATURE}. These materials are represented by the ABX$_3$ structure, where A can be an organic or inorganic constituent (e.g., CH$_3$NH$_3$ or Cs), B is Pb or Sn and the halide is X is a halogen, usually I, Br or Cl. The rich variety of compounds result in a broad range of materials which allow for a fine-tuning of the material properties including the direct optical band gap, mobility, crystal structure or even the isotopic content for nuclear conversion processes. The halide content can be continuously varied, thus fine-tuning the optical properties without compromising the photonics properties \cite{CsPbBr3_halidetuning, lami_halidemixing}. Metal-halide perovskite solar cells reach outstanding photovoltaic efficiency of over 25\%\cite{efficiency-uptodate} thus providing an alternative to silicon-based solar cells. They can also be used as photodetectors\cite{cspbbr3_photodetector2}, X-ray\cite{lami_Xray_Forro}, $\gamma$, and neutron detectors\cite{lami_Xray_Forro}, gas sensors \cite{gazdetektor}, and could be also used in harsh environments including outer space\cite{spaceLAMI, spaceperovskite}.

Organic-hybrid perovskites, such as, e.g., the CH$_3$NH$_3$PbX$_3$ compounds are intensively studied \cite{ScienceReview} due to the added freedom of the organic constituent on the charge transfer, crystalline structure and symmetry \cite{perovskitnapelem}, atom migration, stoichiometric imbalance or enabling of spectroscopic tools including infrared \cite{Perovskite_IR_study} and nuclear magnetic resonance spectroscopies \cite{allignemnt2}. However, one major downside of organic-containing perovskites is their increased sensitivity to environmental effects such as oxygen and humidity \cite{InorganicPerovskiteReview}, whereas fully inorganic perovskites show smaller degradation under ambient conditions. This is an important improvement towards real-world applications.

CsPbBr$_3$ is a particularly compelling material among the inorganic lead-halid perovskites due to a direct optical band gap near the middle of the visible range, a low sensitivity to moisture or air \cite{CsPbBr3_mixing_Cs_stable}, and that Cs is a common constituent in scintillation detectors \cite{CsPbBr3_scintillator}. Furthermore, it is also structurally stable \cite{CsPbBr3_stable}. CsPbBr$_3$ is also a promising candidate for solar cells \cite{CsPbBr3_napelem}, photodetectors\cite{CsPbBr3_detector_PL_Abs, cspbbr3_photodetector2}, LEDs\cite{CsPbBr3_halidetuning}, lasers\cite{CsPbBr3_laser, CsPbBr3_laser2}, and high energy radiation detectors\cite{CsPbBr3_photoconductivity}. 

CsPbBr$_3$ possess orthogonal symmetry below $T=361~\text{K}$, tetragonal phase between $361~\text{K}$ and $403~\text{K}$, and cubic symmetry above $403~\text{K}$ \cite{CsPbBr3_phases}. The band gap of the material is about $2.16~\text{eV}$, which corresponds to approximately a wavelength of $560~\text{nm}$ \cite{CsPbBr3_detector_PL_Abs}, with a slight variation caused by sample preparation and sizes in the nano-scale \cite{CsPbBr3_halidetuning}. The photoluminescent (PL) emission is at $540~\text{nm}$ \cite{CsPbBr3_detector_PL_Abs, CsPbBr3_halidetuning} for bulk single crystals with deviation from this for different sample preparations, sample sizes and multiple PL peaks emerging at low temperatures.\cite{CsPbBr3_halidetuning, CsPbBr3_tobbcsucs1, CsPbBr3_Tfuggo_PL_ABS, CsPbBr3_ketcsucs2, CsPbBr3_TRPL1, CsPbBr3_TRPL3, CsPbBr3_TRPL4}. For most applications, the life-time of photo-generated charge carriers (or $\tauc$) is the vital parameter, which directly determines the photovoltaic or light-emission efficiency. Time-resolved photoluminescence (TRPL) in CsPbBr$_3$ has been studied \cite{CsPbBr3_TRPL1, CsPbBr3_TRPL2, CsPbBr3_TRPL3, CsPbBr3_TRPL4, CsPbBr3_TRPL5} but it is limited to the radiative lifetime. It is thus complementary to the measurement of time-resolved photoconductivity as the latter includes contributions from non-radiative processes, as well. These include, such as the impurity assisted (also known as Shockley-Read-Hall) process and the Auger process. In principle, time-resolved photoconductivity measurements can be performed in CsPbBr$_3$ using a DC technique \cite{CsPbBr3_photoconductivity} or microwave-detected photoconductivity \cite{CsPbBr3_muPCD, CsPbBr3_muPCD2} but such measurements have not been performed for a broad temperature range on single crystals. 

Herein, we report time-resolved microwave-detected photoconductivity (TRMCD) measurements \cite{kunst1, kunst2} in CsPbBr$_3$ for the $20-300~\text{K}$ temperature range. We observe charge-carrier lifetime in excess of $1~\text{ms}$ below $200~\text{K}$ and that even at room temperature $\tauc$ reaches $100~\mu\text{s}$. These values are about an order of magnitude longer than the longest charge-carrier lifetimes found in organic perovskites \cite{BojtorACSPhotonics} and two orders of magnitude longer than those reported in CsPbBr$_3$ previously \cite{CsPbBr3_photoconductivity, CsPbBr3_muPCD, CsPbBr3_muPCD2}. We find compelling evidence from injection-dependent studies, combined with a theoretical modelling that charge trapping is responsible for these long life-times. The presence of charge trapping is supported by observing a similarly long charge-carrier life-time using a continuous-wave excitation near an irradiance with one solar constant. Although a significant trapping of charge carriers may be a limiting factor for photovoltaics and light-emission, it can be advantageous for detector applications. It can also provide an important feedback for improved sample synthesis or it could even find applications as recombination-protective layers in tandem solar cells. \cite{Bencetol_tandem}

\section*{Results and discussion}

\begin{figure}[!ht]
	\centering
	\includegraphics[width=0.98\linewidth]{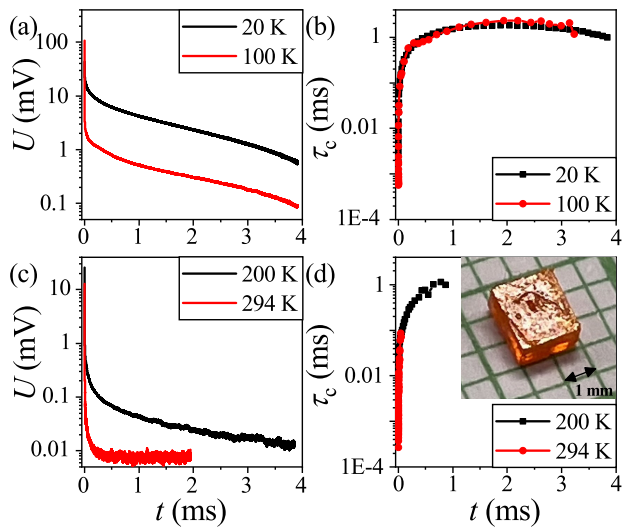}
	\caption{Time-resolved microwave-detected photoconductivity decay curves for CsPbBr$_3$ at different temperatures. (a, c) Raw reflected microwave voltage data, $U(t)$. (b, d) Value of $\tauc$ obtained as discussed in the text. The inset in panel (d) indicates the size of the sample. Note the logarithmic scale for the voltage and $\tauc$ axes. The data shows that $\tauc$ changes significantly during a decay curve and that charge carrier recombination times beyond $1~\text{ms}$ (around $2~\text{ms}$) are present.}
	\label{Fig1_Time_Traces}
\end{figure}

Time-resolved microwave-detected photoconductivity decay measurement results for CsPbBr$_3$ are shown in Fig. \ref{Fig1_Time_Traces}. for temperatures between $20$ and $300$ K as detected with an average irradiance close to one Sun ($100~\text{mW}/\text{cm}^2$). TRMCD is a non-contact, non-destructive method widely used in the semiconductor industry to evaluate the purity and quality of silicon wafers at various steps of preparation. The TRMCD signal is proportional to the charge carrier density as discussed in the Methods section and its time dynamics is dominated by charge recombination processes. It provides the photogenerated charge carrier density, charge carrier mobility, and the recombination time of photoexcited charge carriers, $\tauc$ \cite{Sinton1,sinton2,gyregarami2019ultrafast}. The parameter $\tauc$ determines whether generated charge carriers reach the edge of a solar cell or not, thus greatly influencing the photovoltaic efficiency. 

We employed a Q-switched pulsed laser at $532\ \text{nm}$, $10\ \text{ns}$ pulse width and $200\ \text{Hz}$ repetition rate. Under such conditions, one Sun corresponds to $500~\mu\text{J}/\text{cm}^2$ pulse fluence. The correspondence of the pulsed experiment to continuous-wave conditions is discussed below. Left panels in Fig. \ref{Fig1_Time_Traces}. show the raw time-dependent microwave reflection data, $U(t)$. We could not follow $U(t)$ beyond $4-5\ \text{ms}$ due to the repetition rate being limited to $200\ \text{Hz}$.

We exclude that the observed TRMCD signals are due to sample heating: the presence of the sample itself induces an extra $10\ \text{mV}$ reflected microwave voltage. The total temperature dependence of the "dark" $U(t)$ signal is about $3\ \text{mV}$ in the $20~\text{K}-300~\text{K}$ range. This indicates that the sample in dark conditions is practically an insulator. However, the maximum light-induced $U(t)$ is above $10\ \text{mV}$ for one Sun irradiance and it is in fact about $100\ \text{mV}$ at low temperatures. Therefore such a large reflected microwave signal could not be achieved simply by sample heating. This also means that all the reported time-dependent results are genuine and are related to the recombination dynamics of free charge carriers. We also reproduced the same results on six different samples, including those which were freshly cleaved and also samples with very different sample sizes and shapes and we observed no significant variation among the samples.

$U(t)$ is then evaluated to yield the apparent charge-carrier recombination time\cite{muPCD_lauer, tau_dn_derivalt_calculation}, $\tauc$, from the rate equation as (see Methods for more details):
\begin{equation}\label{tau_c_formula_0}
	\tauc = -\frac{U(t)}{\frac{\partial U(t)}{\partial t}}.
\end{equation}

The result is shown on the right panels of Figure \ref{Fig1_Time_Traces}., i.e. $\tauc$ as a function of time. The data points are smoothed (details described in the Methods section) to reduce the apparent noise in the numerical derivative.

The important observations are that i) $\tauc$ is short (down to $1\ \mu\text{s}$) immediately after the pulse, and ii) it steadily increases and reaches recombination times as long as $1$ ms long after the pulse, except for the room temperature data. While the presence of a long $\tauc$ at low temperatures is apparent in our work, we discuss why it has not been observed around room temperature in previous studies \cite{CsPbBr3_muPCD, CsPbBr3_muPCD2}, although a $\tauc$ up to $100~\mu\text{s}$ is also present therein. One possibility is that sample purity is very different in our work. We rather believe that our high sensitivity setup (details given in the Methods section), which uses a coplanar waveguide, a low-noise microwave amplifier, a balanced microwave bridge, and a state-of-art mixer and extensive averaging, is key for this observation.

Time dependence of the charge-carrier relaxation time is not unexpected. In Si, it is well known that $\tauc$ depends on the excess charge-carrier concentration, $\Delta n$\cite{muPCD_lauer,Feriek_QSSPCD_tau_dn, dn_tau_connection1} due to an interplay between several recombination mechanisms, including the Schockley-Read-Hall (SRH), the radiative, and Auger recombinations.\cite{SRH_Auger_rad, Hall_alapmu} Individually, these recombination methods depend differently on $\Delta n$: the SRH recombination is mostly independent of $\Delta n$ but the resulting $\tauc$ changes between two values for low and high injection levels. The two- and three-particle radiative and Auger processes give $\tauc$ being as a linear and quadratic function of $\Delta n$, respectively.

\begin{figure*}[!ht]
	\centering
	\includegraphics[width=0.98\linewidth]{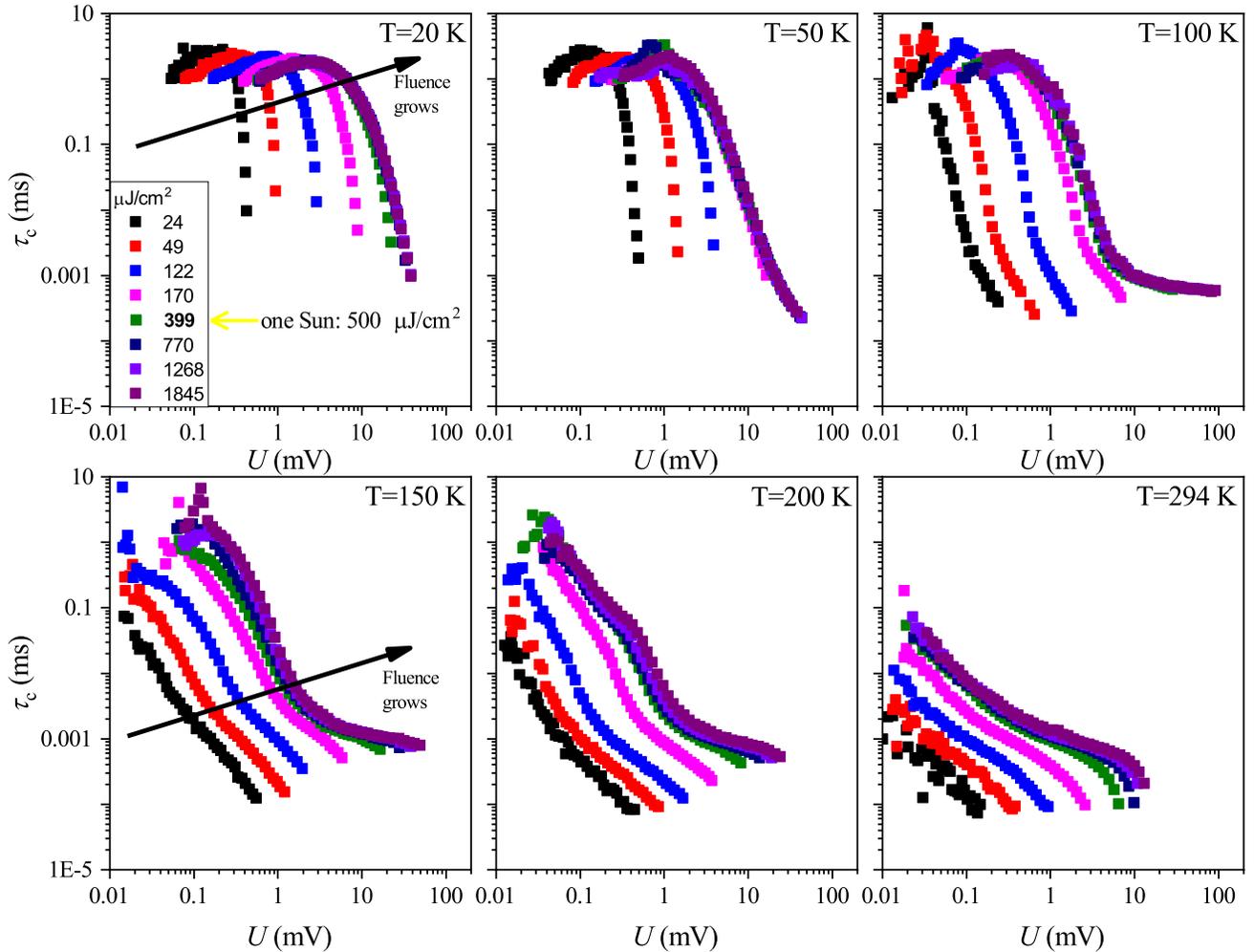}
	\caption{Charge-carrier recombination times for various temperatures and exciting laser pulse energies shown as a function of the reflected microwave voltage. The latter quantity is a measure of the excess charge-carrier concentration. Yellow arrow indicates the laser energy value which corresponds to about one Sun ($500~\mu\text{J}/\text{cm}^2$ with our $200$ Hz repetition rate). We observe a gradual shifting of the curves toward higher detected voltages with increasing laser pulse energy, while their shape does not change.}
	\label{Fig2_Proof_Trapping}
\end{figure*}

The underlying recombination mechanisms can be also studied by varying the exciting pulse energies. Fig. \ref{Fig2_Proof_Trapping}. shows the $\tauc$ data as a function of the reflected microwave voltage. The latter quantity is proportional to $\Delta n$ under reasonable assumptions as discussed in the Methods section. However, a direct conversion of $U$ to $\Delta n$ is the topic of a future work. The data is shown in Fig. \ref{Fig2_Proof_Trapping} for six different temperatures and for eight different laser pulse energies, one of which corresponds roughly to an average power of one Sun ($0.1$ mW/$\text{cm}^2$). 

The behavior of the $\tauc$ curves is somewhat unusual as e.g., for Si, similar curves fall on one another, and the different laser energies, in fact, map different areas of the same curve. In contrast, for CsPbBr$_3$ the curves shift progressively toward higher reflected microwave voltages, while their underlying shapes change little. As we show below, this is characteristic to a trap-dominated charge-carrier recombination mechanism. In fact, there is mounting evidence for the presence or even for the dominance of charge carrier trapping in metal-halide perovskites both for organic \cite{Organic_Halide_trapping} and inorganic \cite{cspbbr3_trappak_meres} compounds and also the theoretical description of this effect has been given in Ref. \onlinecite{cspbbr3_trapping_elmelet}.

\begin{figure}[!ht]
	\centering
	\includegraphics[width=0.98\linewidth]{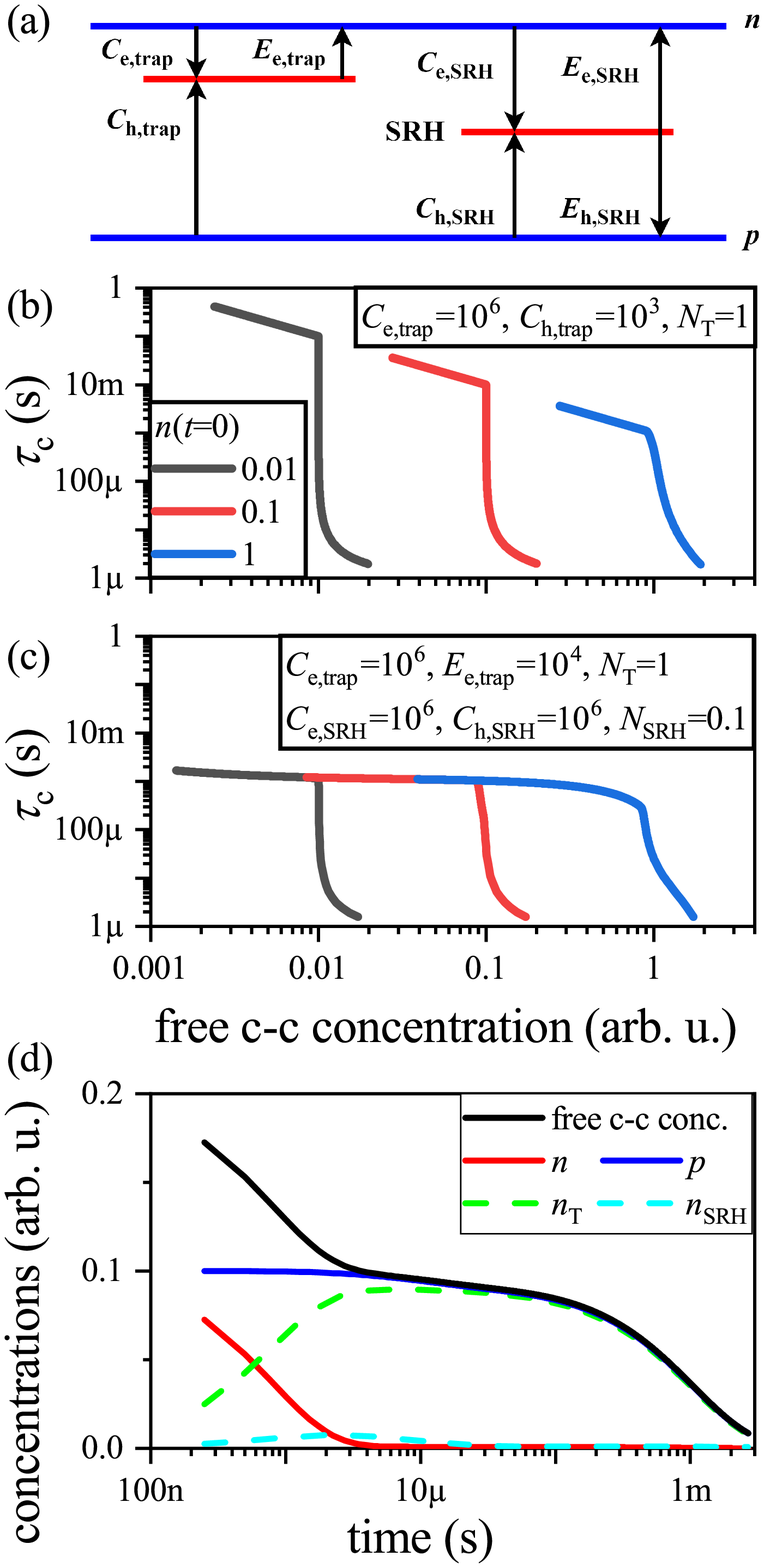}
	\caption{Band model of the charge-carrier dynamics in CsPbBr$_3$ (a) and the resulting charge-carrier concentration dependent recombination times for various initial injection levels: (b) recombination with a trap level only, (c) recombination in the presence of a trap+SRH states. The values of the non-zero transition rates are shown. (d) Time dependence of the individual charge states for the trap+SRH case with $n(t=0)=0.1$.}
	\label{model}
\end{figure}

We attempt to describe the charge-carrier dynamics in CsPbBr$_3$ with a minimal model. The model is depicted in Fig. \ref{model}a.: it involves the capture of electrons with density $n$ in the conduction band (CB) at a rate of $C_\text{e,trap}$ and their emission into it at a rate of $E_\text{e,trap}$ as well as capture of a hole into the trap from the valence band (VB) at a rate of $C_\text{h,trap}$. The latter process is in fact the recombination and physically corresponds to the emission of an electron into the VB. The density of trap states is denoted by $N_\text{T}$ and the density of occupied trap states by $n_\text{T}$. The corresponding set of rate equations reads as:
\begin{align} \label{Diff_eq_ver1}
    \frac{\text d n}{\text d t}=& -C_\text{e,trap}\left( N_\text{T}- n_\text{T}\right) n+E_\text{e,trap} n_\text{T}, \nonumber \\
    \frac{\text d n_\text{T}}{\text d t}=&+C_\text{e,trap}\left( N_\text{T}- n_\text{T}\right)n-E_\text{e,trap} n_\text{T} \nonumber \\ &-C_\text{h,trap}n_\text{T} p, \nonumber \\
    \frac{\text d p}{\text d t}=&-C_\text{h,trap}n_\text{T} p,
\end{align}
where charge neutrality dictates that ${{p=n+n_\text{T}}}$. We note that the dimensions of emission and capture rates (or cross sections) differ but for simplicity we consider both to be dimensionless as the considered charge levels are also around unity. We also note that this model involves the trapping of electrons but it could naturally be replaced by considering trapping of holes instead.

Solution of Eq. \eqref{Diff_eq_ver1} yields the charge-carrier concentrations as well as the effective $\tauc$ as a function of time and is shown in Fig. \ref{model}b. We fixed the different capture/emission rates as given in the figure and varied the initial charge injections denoted as $n(t=0)$ (equal to $p(t=0)$). The main feature is that the $\tauc$ \textit{versus} time curves shift with varying $n(t=0)$ as expected for the trapping mechanism, the two main lifetimes determined by $C_\text{e,trap}=10^6$ and $C_\text{h,trap}=10^3$, while keeping $C_\text{e,trap}=0$, whose actual value does not significantly affect the curves. While this simple model explains some features in the measured TRMCD traces, it fails in explaining the constant $\tauc$ behavior after the trap state is saturated. A constant $\tauc$ value, i.e., independent of $\Delta n$, rather resembles the normal Schockley-Read-Hall (SRH) recombination (also known as trap-assisted recombination).

We therefore amended our model with the presence of an SRH centrum which is also shown schematically in Fig. \ref{model}c. In this model, the recombination is dominated by the SRH process through SRH states with density $N_{\text{SRH}}$ (and occupation $n_{\text{SRH}}$), it is therefore insensitive of $C_\text{h,trap}$, which is set to $0$. However, the emission of trapped electrons becomes relevant, which is a thermally activated process and we set the corresponding rate to $E_\text{e,trap}=10^4$. The model also includes the electron and hole capture rates into the SRH state (with rates $C_\text{e,SRH}$ and $C_\text{h,SRH}$, respectively) and also their emissions ($E_\text{e,SRH}$ and $E_\text{h,SRH}$ being the respective rates). The latter rates are thermally activated, their values are thus set to $0$ in our model. We found that the constant $\tauc$ behavior can only be explained when $N_\text{SRH}\ll N_\text{trap}$, therefore, we use $N_\text{SRH}=0.1$ and $N_\text{trap}=1$ in our calculations.

A set of rate equations, similar to that in Eq. \eqref{Diff_eq_ver1} can be established for the amended model and solved with the above parameters, and the result is shown in Fig. \ref{model}d. Clearly, both the trapping dominated rapid charge depletion as well as the saturated $\tauc$ behavior is well explained with this simple model. Fig. \ref{model}d. shows the time dependence for the individual charge state components for the trap+SRH model when $n(t=0)=0.1$. Much as the model explains the low temperature behavior, it does not account for the temperature dependence of $\tauc$, which probably requires the inclusion of the electron and hole emission parameters, too. A more detailed theoretical study is beyond our present scope, nevertheless, we believe that the detailed temperature and charge injection dependent data may serve as useful input for a full description.

\begin{figure}[!ht]
	\centering
	\includegraphics[width=0.98\linewidth]{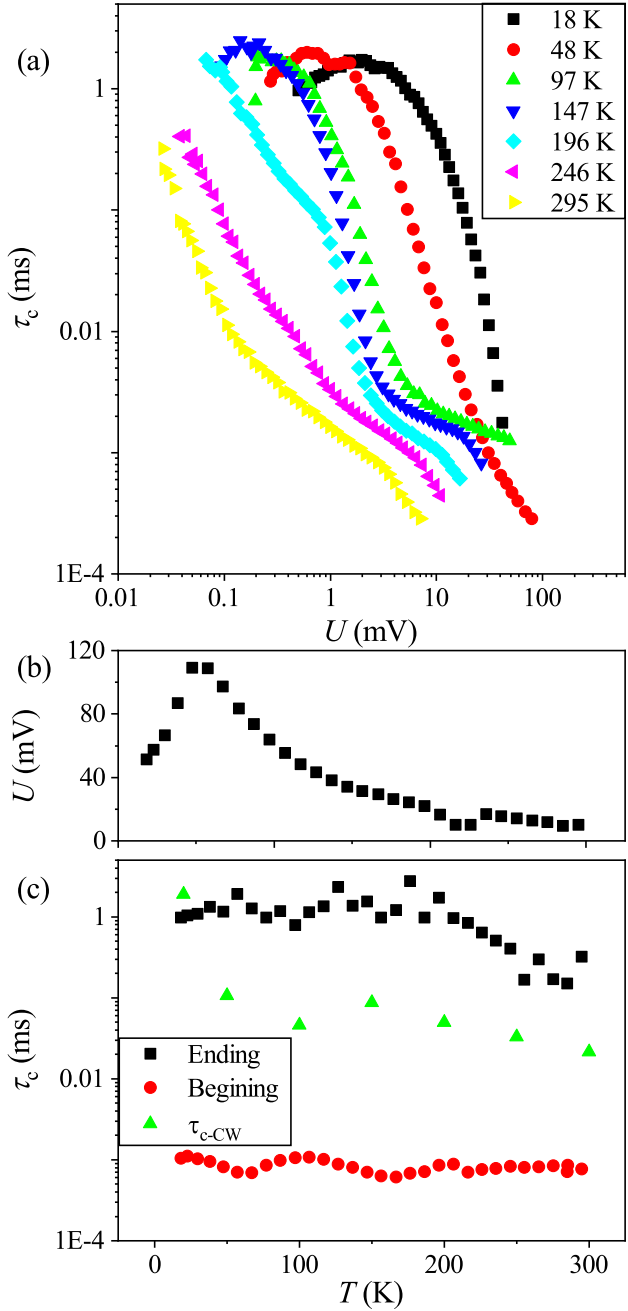}
	\caption{(a) Temperature dependence of the charge carrier recombination curves. (b) The initial amplitude of the TRMCD data ($U$). (c) The shortest and longest $\tauc$ times. The effective $\tau_\text{c,CW}$, as explained in the text, is also given in (c).}
	\label{Fig4_tau_temp_ampl}
\end{figure}

%\subsection*{Analysis of the CW measurements}
Beyond characterizing the recombination dynamics after short pulse illumination, the material’s response to steady-state illumination is also important for the assessment of using it for photovoltaic devices. Therefore we also performed continuous-wave (CW) measurements with a total irradiance similar to one Sun. This in principle provides additional information about the mechanism of the charge carrier recombination. Typically, steady-state lifetime measurements requires the precise value of electron and hole mobilities to convert the measured signal to $\Delta n$. In our cases, this information is not available. However, to overcome this obstacle, we used a tricky way to determine steady-state lifetime, using the amplitude of the TRMCD measurement for this conversion. In the following, we compare measurements with the pulsed and CW lasers operating with the same average power. The energy of the pulsed laser is compressed into $t_\text{pulse}=10~\text{ns}$ long pulses (with a repetition rate of $f_\text{pulse}=200~\text{Hz}$). We denote the charge carrier generation by the pulsed and CW lasers by $G_\text{pulse}$ and $G_\text{CW}$, respectively. When the average powers are the same, the two are related by: $G_\text{pulse}=\frac{1}{t_\text{pulse} f_\text{pulse}}\cdot G_\text{CW}$. As discussed in the Methods section, the amount of excess charge carriers following a pulse is $\Delta n_\text{pulse}=G_\text{pulse}\cdot t_\text{pulse}$ and for the CW case it is $\Delta n_\text{CW}=G_\text{CW}\cdot \tau_\text{c,CW}$. Here, we introduced the notation $\tau_\text{c,CW}$, which can be considered as a proportionality constant between $\Delta n_\text{CW}$ and $G_\text{CW}$ and is related to an effective lifetime of charge carriers under CW illumination conditions. Since the laser pulse is much shorter than the observed decay times, we can neglect the recombination during the rapid exaction.

Assuming that the detected microwave voltage is not very different in the two cases, thus we are in a regime where $U\propto \Delta n$ is valid, we obtain for the ratio of the measured voltage following a pulse and that in the CW experiment as:
\begin{equation}
    \frac{U_\text{CW}}{U_\text{pulse}}=f_\text{pulse}\cdot \tau_\text{c,CW}.
\end{equation}

This equation allows us to directly determine $\tau_\text{c,CW}$ from the data and compare it to the results of the TRMCD measurements. For the latter, $\tauc$ changes with time and $\Delta n$, we can thus define it for the "beginning" and "ending" of a time trace from the individual curves at different temperatures as show in Fig. \ref{Fig4_tau_temp_ampl}a. Remarkably, the $\tau_\text{c,CW}$ values fall in between the two types of lifetime data as shown in Fig. \ref{Fig4_tau_temp_ampl}c. and are above $100~\mu\text{s}$ throughout. In fact finding such a long $\tau_\text{c,CW}$ strongly supports the charge trapping scenario: were $\tauc$ primarily dependent on $\Delta n$, such as in Si, we should have found strongly different tau values.

Concerning the temperature-dependence amplitude of the TRMCD curves, we observe a maximum around $45-55$ K and a decrease around $220~\text{K}$ as shown Fig. \ref{Fig4_tau_temp_ampl}b. Both effects were reproduced on a number of samples and for several temperature runs including various excitation levels. The maximum means an increase in photoconductance and/or photogeneration efficiency but we are aware of no known structural or electronic changes which may be associated with this. There exist reports of an anomalous behavior in CsPbBr$_3$ around $220$ K of the dielectric function \cite{CsPbBr3_effectat220K} and also the thermal emission from trap levels \cite{cspbbr3_trappak_meres}, supported by a theoretical description \cite{cspbbr3_trapping_elmelet} based on Pb-Br antisites, which may be also reflected in our TRMCD data.

\section*{Conclusions}

In summary, we conducted temperature and power-dependent time-resolved microwave photoconductivity decay measurements on CsPbBr$_3$ single crystals in the temperature range of $20-300~\text{K}$ around the average excitation of one Sun. We observe ultra-long, beyond $1~\text{ms}$ charge carrier lifetime in CsPbBr$_3$ single crystals below $200~\text{K}$. The injection-dependent TRMCD measurements indicate the observed ultra-long effect to be caused by charge carriers being trapped in shallow traps previously reported in CsPbBr$_3$. The measured ultra-long lifetime is consistent with the results obtained for CW measurements conducted at the same average excitation power as the TRMCD measurements. While the trapping of charge carriers causes one type of carrier to be immobile, thus is a limiting factor for direct photovoltaic applications, the observed ultra-long lifetimes can still be useful in other applications including sensitive photodetection for other wavelengths, for example, X-rays, light emission, quantum-memory storage or special photovoltaic structures such as tandem cells.

\section*{Methods and tools}

\subsection*{Sample preparation}

\subsubsection*{Chemicals and reagents}

PbBr$_2$ ($98+\%$) and DMSO ($99.8+\%$) were purchased from Thermo Scientific, CsBr ($99\%$) was purchased from Alfa Aesar, and DMF ($99\%$) was purchased from Fisher Scientific.

\subsubsection*{Preparation of the precursor solution}

The precursor solution of CsPbBr$_3$ was prepared by dissolving PbBr$_2$ and CsBr in $2:1$ molar ratio in dymethylsulfoxide (DMSO). The precursor solution is heated to $80~^{\circ}\text{C}$ to accelerate the dissolution of the salts. A transparent solution is obtained after $10~\text{h}$ of stirring, after which it is filtered. This preparation of the precursor solution is a modified version of the method published by Dirin \emph{et al}. \cite{CsPbBr3_sampleprepare3}.

\subsubsection*{Growth of CsPbBr$_3$ single crystals}

The crystals are grown using the inverse temperature crystal growth technique (ITC) initially proposed by Saidaminov \emph{et al}. \cite{CsPbBr3_sampleprepare1, CsPbBr3_sampleprepare2}. The solubility of the perovskite is decreasing by increasing the temperature from $80~^{\circ}\text{C}$ to $120~^{\circ}\text{C}$ in DMSO. Crystals are then collected and employed as seed crystals for a further growth step. They are placed in a fresh precursor solution whose temperature is slowly heated up to $110~^{\circ}\text{C}$. The resulting single crystal is harvested, rinsed with dimethylformamide (DMF) solvent and dried.

\subsection*{The microwave-detected photoconductivity setup}

\begin{figure}[!ht]
	\centering
	\includegraphics[width=0.98\linewidth]{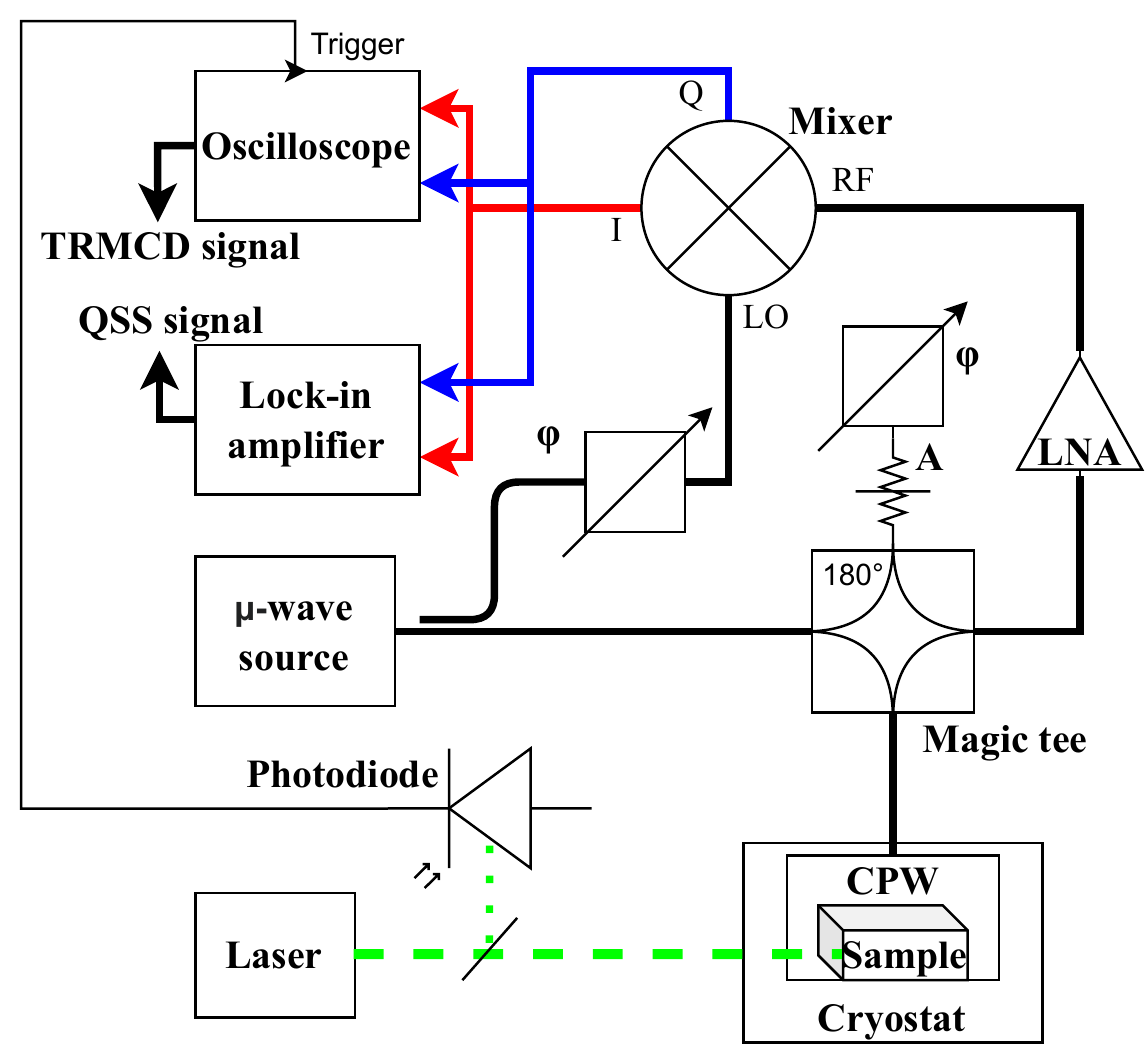}
	\caption{Schematics of the experimental setup, showing both the CW and pulsed laser, whose use is optional, along with the respective use of a lock-in amplifier for CW and an oscilloscope for the pulsed measurements. A photodiode is used for triggering in the pulsed experiment.}
	\label{block}
\end{figure}

We used a pulsed laser (NL201-2.5k-SH-mot, Ekspla) with an adjustable repetition rate, pulse energy, and wavelength (selectable between $632$ and $1064$ nm). We used excitation at $532~\text{nm}$, the repetition rate was $200~\text{Hz}$. Due to the wavelength of the laser, we were able to excite the material near its absorption edge and due to the repetition rate, we had the opportunity to observe the slow recombination process found at low temperatures. We used a high-speed Tektronix MDO3024 oscilloscope to record the time traces of the relaxation. A photodetector (DET36A/M, Thorlabs), placed after a beam sampler provided the trigger signal for the oscilloscope. 

The microwave reflectometry setup is also shown in Figure \ref{block}. The microwave source (MKU LO 8-13 PLL, Kühne GmbH) was set to $10~\text{GHz}$. We used a hybrid coupler (R433721, Microonde) to split the signal in two with the purpose of providing an LO signal for the IQ mixer. We used a magic tee with a reference arm equipped with phase and amplitude tuning to cancel out the DC reflection coming from the sample leaving only the AC component due to photoconductivity. This way the saturation of the mixer can be avoided and the oscilloscope is kept at the highest possible digital resolution. We included isolators to shield the microwave source from any signals going toward its output and DC blocks before the mixer. We used an IQ0618LXP mixer (Marki Microwaves) with the signal reflected from the sample connected to the RF arm and the reference signal connected to the LO arm. We used a JaniLab Inc. low-noise amplifier between the magic tee and the IQ mixer. The sample is placed on a coplanar waveguide (CPW) inside a cryostat (M-22, CTI-CRYOGENICS). The temperature is variable between $10\ \text{K}$ and room temperature.

Coplanar waveguides are planar structures consisting of a ceramic layer that has conducting areas on it divided by gaps. These instruments are ideal for the microwave regime but can be used in a wide range of wavelengths. The coplanar waveguide we used is a conductor-backed coplanar waveguide (CBCPW). This means that the signal travels in a central strip while the two strips beside it and the back side of the CPW act as the ground. The front and backside ground surfaces are connected with via-holes through the ceramic layer, holes between the front and backside coated with the same conductive material as the surfaces. The magnetic and electric field of a CPW \cite{CPWterek} is ideal for the measurements presented herein since the magnetic field is greatest and most homogeneous at the gap between the signal and ground strip on the front side resulting in high signal levels when the sample is placed at this point. By placing the sample on the CPW located on the cold finger of the cryostat we can create a measurement setup where the CPW used as an antenna also functions as the cooling pad for the sample.

With this setup, we directly measure the microwave signal reflected by the sample. Due to the DC signal reduction realized by the reference arm of the magic tee we can measure the effect of excitation without the DC reflection coming from the sample. Using a phase shifter, placed between the source and the LO arm of the mixer, the phase difference between the RF and LO arm is also tunable thus providing a way to change the ratio of signal between the I and Q arm of the output.

\subsection*{Analysis of the coplanar waveguide based measurements}

Strictly speaking, our experimental setup measures the reflected microwave voltage, both in- and out-of-phase due to the IQ mixer based phase sensitive detection. This inevitably contains some residual DC background. The exciting, $U_{\text{exciting}}$ and reflected, $U_{\text{reflected}}$ voltages are related to each other with the reflection constant, $\Gamma$ also known as the $S_{11}$ parameter, which is generally constant, due to a possible phase shift in the reflected voltage. The reflection occurs as the presence of the sample on the CPW perturbs locally its wave-impedance, $Z_0=50~\Omega$ \cite{pozar}, giving rise to $Z_{\text{perturbed}}$. This leads to the well-known relationship between the reflection coefficient and the perturbed impedance:
\begin{equation}
	\Gamma=\frac{U_{\text{reflected}}}{U_{\text{exciting}}}=\frac{Z_{\text{perturbed}}-Z_0}{Z_{\text{perturbed}}+Z_0}.
\end{equation}

The perturbation occurs due to the so-called surface impedance of the sample, $Z_{\text{s}}$. The concept of the surface impedance is a convenient way to describe the high-frequency properties (radiofrequency or microwave) of a material but mathematically it is equivalent to using the concept of the complex refractive index in optics. Its explicit form at an angular frequency of $\omega$ reads:
\begin{equation}
	Z_{\text{s}}=\sqrt{\frac{\mathrm{i}\omega\mu}{\sigma+\mathrm{i}\omega\epsilon}},
\end{equation}
which contains the effects of magnetism (through the permeability, $\mu$), dielectrics (through the permittivity, $\epsilon$), and conductivity (through $\sigma$). 

In the presence of a sample, the resulting perturbed CPW impedance reads: $Z_{\text{perturbed}}=Z_0+\eta Z_{\text{s}}$. Herein, the $\eta$ factor depends on the sample size and it describes how the sample covers the gap between the grounding and the center conductor of the CPW. Its accurate value is sample dependent and is unknown in our study.

Microwave-dependent photoconductivity measurements generally assume \cite{muPCD_lauer} that the reflected voltage and the sample conductivity are linear or at least the relationship can be linearized for a reasonable $\sigma$ range. Under these assumptions, we obtain that the additional microwave reflection due to the presence of the sample, $\Delta \Gamma$ reads:
\begin{equation}
	\Delta \Gamma \approx\Gamma_0\cdot \frac{\Delta \sigma}{\sigma_0}\propto \frac{\Delta n}{n_0},
\end{equation}
where $\Delta n$ is the excess charge carrier concentration, $n_0$, $\Gamma_0$ and $\sigma_0$ are the charge carrier concentration, reflectivity coefficient, and conductivity for the sample without illumination. This equation also means that the relationship between the measured reflected microwave voltage, $U_{\text{reflected}}$ (denoted as $U$ for simplicity in the following) and the excess charge carrier concentration $\Delta n$ is linear, i.e. $U=C\cdot \Delta n$.

Below, we discuss a more generic case, i.e., when the relationship is not necessarily linear: $U=f\Delta n$. Here, $f$ is an unknown, analytic function that depends on several experimental factors, the details of the CPW, on the microwave frequency, etc. A careful analysis shows that in this case, no linear assumptions are required to obtain the time-dependent $\Delta n$ values in the experiment as well as the corresponding charge-carrier recombination time.

The time dynamics of the excess charge carrier concentration is governed by the rate equation:
\begin{equation} \label{cc_diff_eq}
	\frac{\partial \Delta n(t)}{\partial t}=-\frac{\Delta n}{\tauc} +G,
\end{equation}
where $G$ is the rate of photogeneration with dimensions of charge per volume per unit time. Albeit Eq. \eqref{cc_diff_eq} appears to be a simple linear differential equation, $\tauc$ is not necessarily a constant as it may depend on $\Delta n$ or $t$. Eq. \eqref{cc_diff_eq} can be treated in different regimes, a) including during a laser pulse, b) for a steady state, and c) the relaxation of the generated charge carriers following a laser pulse. 

During a laser pulse, when the pulse duration is shorter than the typical value of $\tauc$, $\Delta n(t)$ grows linearly with time and after a duration of $t$ the amount of photogenerated charge carriers simply reads $\Delta n=G_{\text{pulse}}t$. Here we introduced $G_{\text{pulse}}$ for the photogeneration rate during the pulse.

When the material is irradiated with a CW laser, i.e., for the steady-state state, we obtain $\Delta n_{\text{CW}}=G_{\text{CW}}t$, where we introduced $G_{\text{CW}}$ for the photogeneration rate by the CW laser.

Following a pulse, $\tauc$ can be obtained from Eq. \eqref{cc_diff_eq} as:
\begin{equation}
	\tauc=-\frac{\Delta n(t)}{\frac{\partial \Delta n(t)}{\partial t}},
\label{tau_c_formula_Delta_n}
\end{equation}
even when $\tauc$ depends on the excess charge carrier concentration (which is the case for most relaxation processes including the Auger, and radiative recombination, and to some extent for the SRH process). As a result, $\tauc$ varies during a time trace itself, since $\Delta n$ also changes.

When the above-mentioned linear relationship between $U$ and $\Delta n$ holds, the value of $\tauc$ can be obtained directly from the time traces of the microwave reflection voltage data ($U(t)$) using Eq. \eqref{tau_c_formula_Delta_n} as:
\begin{equation}
	\tauc=-\frac{U(t)}{\frac{\partial U(t)}{\partial t}}.
\label{tau_c_formula}
\end{equation}

We use Eq. \eqref{tau_c_formula} throughout herein to obtain the results on $\tauc$. Please note, that the method presented above provides the accurate carrier lifetime in the case of ordinary recombination process, if carrier trapping phenomena is not as significant as in our case. Since $\Delta n >> \Delta p$, the signal in steady state is dominantly originating from excess electrons, and therefore the evaluated lifetime corresponds to the mean time of electrons conductive state before getting recombined. Since the obtained values are consistent with the observed decay times, it can be concluded the method is useful to characterize the recombination properties of excess electrons in steady-state condition.

As we measure a lot more time points than shown in the relevant figures and also the numerical derivative is very noisy, we employ an adaptive smoothing on the data. First, the maximum value of a time trace is obtained (among the first few data points in the time trace following a pulse) and we also calculate the minimum value of a time trace, which is obtained long after the pulse by averaging the noise. Then, the vertical values of the data points are segmented into $n$ bins ($n$ being typically 50-100) where the bin values are distributed logarithmically between the minimum and maximum data points. This essentially reflects the exponentially decaying nature of the time traces. Data points in a given bin are averaged out which results in a single point per bin. This method results in an essentially adaptive smoothing of the otherwise noisy data. 

\section*{Acknowledgements}
The Authors are indebted to R.~Ga\'{a}l for the help with the instrument. Work supported by the National Research, Development and Innovation Office of Hungary (NKFIH), and by the Ministry of Culture and Innovation Grants Nr. K137852, 2022-2.1.1-NL-2022-00004, 2019-2.1.7-ERA-NET-2021-00028 (V4-Japan Grant, BGapEng), C1010858 (KDP-2020) and C1530638 (KDP-2021).

%\bibliography{LAMI_liter}

\begin{mcitethebibliography}{57}
	\providecommand*\natexlab[1]{#1}
	\providecommand*\mciteSetBstSublistMode[1]{}
	\providecommand*\mciteSetBstMaxWidthForm[2]{}
	\providecommand*\mciteBstWouldAddEndPuncttrue
	{\def\EndOfBibitem{\unskip.}}
	\providecommand*\mciteBstWouldAddEndPunctfalse
	{\let\EndOfBibitem\relax}
	\providecommand*\mciteSetBstMidEndSepPunct[3]{}
	\providecommand*\mciteSetBstSublistLabelBeginEnd[3]{}
	\providecommand*\EndOfBibitem{}
	\mciteSetBstSublistMode{f}
	\mciteSetBstMaxWidthForm{subitem}{(\alph{mcitesubitemcount})}
	\mciteSetBstSublistLabelBeginEnd
	{\mcitemaxwidthsubitemform\space}
	{\relax}
	{\relax}
	
	\bibitem[Correa-Baena \latin{et~al.}(2017)Correa-Baena, Saliba, Buonassisi,
	Graetzel, Abate, Tress, and Hagfeldt]{ScienceReview}
	Correa-Baena,~J.-P.; Saliba,~M.; Buonassisi,~T.; Graetzel,~M.; Abate,~A.;
	Tress,~W.; Hagfeldt,~A. Promises and challenges of perovskite solar cells.
	\emph{Science} \textbf{2017}, \emph{358}, 739--744\relax
	\mciteBstWouldAddEndPuncttrue
	\mciteSetBstMidEndSepPunct{\mcitedefaultmidpunct}
	{\mcitedefaultendpunct}{\mcitedefaultseppunct}\relax
	\EndOfBibitem
	\bibitem[Huang \latin{et~al.}(2017)Huang, Yuan, Shao, and Yan]{efficiency}
	Huang,~J.; Yuan,~Y.; Shao,~Y.; Yan,~Y. Understanding the physical properties of
	hybrid perovskites for photovoltaic applications. \emph{Nat. Rev. Mater.}
	\textbf{2017}, \emph{2}, 17042\relax
	\mciteBstWouldAddEndPuncttrue
	\mciteSetBstMidEndSepPunct{\mcitedefaultmidpunct}
	{\mcitedefaultendpunct}{\mcitedefaultseppunct}\relax
	\EndOfBibitem
	\bibitem[Lin \latin{et~al.}(2018)Lin, Xing, Quan, de~Arquer, Gong, Lu, Xie,
	Zhao, Zhang, Yan, Li, Liu, Lu, Kirman, Sargent, Xiong, and
	Wei]{PerovskiteLED_NATURE}
	Lin,~K. \latin{et~al.}  Perovskite light-emitting diodes with external quantum
	efficiency exceeding 20 per cent. \emph{Nature} \textbf{2018}, \emph{562},
	245+\relax
	\mciteBstWouldAddEndPuncttrue
	\mciteSetBstMidEndSepPunct{\mcitedefaultmidpunct}
	{\mcitedefaultendpunct}{\mcitedefaultseppunct}\relax
	\EndOfBibitem
	\bibitem[Protesescu \latin{et~al.}(2015)Protesescu, Yakunin, Bodnarchuk, Krieg,
	Caputo, Hendon, Yang, Walsh, and Kovalenko]{CsPbBr3_halidetuning}
	Protesescu,~L.; Yakunin,~S.; Bodnarchuk,~M.~I.; Krieg,~F.; Caputo,~R.;
	Hendon,~C.~H.; Yang,~R.~X.; Walsh,~A.; Kovalenko,~M.~V. Nanocrystals of
	Cesium Lead Halide Perovskites (CsPbX3, X = Cl, Br, and I): Novel
	Optoelectronic Materials Showing Bright Emission with Wide Color Gamut.
	\emph{Nano Lett.} \textbf{2015}, \emph{15}, 3692--3696\relax
	\mciteBstWouldAddEndPuncttrue
	\mciteSetBstMidEndSepPunct{\mcitedefaultmidpunct}
	{\mcitedefaultendpunct}{\mcitedefaultseppunct}\relax
	\EndOfBibitem
	\bibitem[Zhang \latin{et~al.}(2015)Zhang, Zhong, Chen, Wu, Hu, Huang, Han, Zou,
	and Dong]{lami_halidemixing}
	Zhang,~F.; Zhong,~H.; Chen,~C.; Wu,~X.-g.; Hu,~X.; Huang,~H.; Han,~J.; Zou,~B.;
	Dong,~Y. {Brightly Luminescent and Color-Tunable Colloidal
		CH$_3$NH$_3$PbX$_3$ (X = Br, I, Cl) Quantum Dots: Potential Alternatives for
		Display Technology}. \emph{ACS Nano} \textbf{2015}, \emph{9},
	4533--4542\relax
	\mciteBstWouldAddEndPuncttrue
	\mciteSetBstMidEndSepPunct{\mcitedefaultmidpunct}
	{\mcitedefaultendpunct}{\mcitedefaultseppunct}\relax
	\EndOfBibitem
	\bibitem[Green \latin{et~al.}()Green, Dunlop, Hohl-Ebinger, Yoshita, Kopidakis,
	and Hao]{efficiency-uptodate}
	Green,~M.~A.; Dunlop,~E.~D.; Hohl-Ebinger,~J.; Yoshita,~M.; Kopidakis,~N.;
	Hao,~X. Solar cell efficiency tables (version 59). \emph{Prog. Photovolt.}
	\emph{30}, 3--12\relax
	\mciteBstWouldAddEndPuncttrue
	\mciteSetBstMidEndSepPunct{\mcitedefaultmidpunct}
	{\mcitedefaultendpunct}{\mcitedefaultseppunct}\relax
	\EndOfBibitem
	\bibitem[Shamsi \latin{et~al.}(2017)Shamsi, Rastogi, Caligiuri, Abdelhady,
	Spirito, Manna, and Krahne]{cspbbr3_photodetector2}
	Shamsi,~J.; Rastogi,~P.; Caligiuri,~V.; Abdelhady,~A.~L.; Spirito,~D.;
	Manna,~L.; Krahne,~R. Bright-Emitting Perovskite Films by Large-Scale
	Synthesis and Photoinduced Solid-State Transformation of CsPbBr$_3$
	Nanoplatelets. \emph{ACS Nano} \textbf{2017}, \emph{11}, 10206--10213\relax
	\mciteBstWouldAddEndPuncttrue
	\mciteSetBstMidEndSepPunct{\mcitedefaultmidpunct}
	{\mcitedefaultendpunct}{\mcitedefaultseppunct}\relax
	\EndOfBibitem
	\bibitem[Náfrádi \latin{et~al.}(2015)Náfrádi, Náfrádi, Forró, and
	Horváth]{lami_Xray_Forro}
	Náfrádi,~B.; Náfrádi,~G.; Forró,~L.; Horváth,~E. Methylammonium Lead
	Iodide for Efficient X-ray Energy Conversion. \emph{J. Phys. Chem. C}
	\textbf{2015}, \emph{119}, 25204--25208\relax
	\mciteBstWouldAddEndPuncttrue
	\mciteSetBstMidEndSepPunct{\mcitedefaultmidpunct}
	{\mcitedefaultendpunct}{\mcitedefaultseppunct}\relax
	\EndOfBibitem
	\bibitem[Mantulnikovs \latin{et~al.}(2018)Mantulnikovs, Glushkova, Matus,
	Ćirić, Kollár, Forró, Horváth, and Sienkiewicz]{gazdetektor}
	Mantulnikovs,~K.; Glushkova,~A.; Matus,~P.; Ćirić,~L.; Kollár,~M.;
	Forró,~L.; Horváth,~E.; Sienkiewicz,~A. Morphology and Photoluminescence of
	$\text{CH}_3\text{NH}_3\text{PbI}_3$ Deposits on Nonplanar, Strongly Curved
	Substrates. \emph{ACS Photonics} \textbf{2018}, \emph{5}, 1476--1485\relax
	\mciteBstWouldAddEndPuncttrue
	\mciteSetBstMidEndSepPunct{\mcitedefaultmidpunct}
	{\mcitedefaultendpunct}{\mcitedefaultseppunct}\relax
	\EndOfBibitem
	\bibitem[Ho-Baillie \latin{et~al.}(2022)Ho-Baillie, Sullivan, Bannerman,
	Talathi, Bing, Tang, Xu, Bhattacharyya, Cairns, and McKenzie]{spaceLAMI}
	Ho-Baillie,~A. W.~Y.; Sullivan,~H. G.~J.; Bannerman,~T.~A.; Talathi,~H.~P.;
	Bing,~J.; Tang,~S.; Xu,~A.; Bhattacharyya,~D.; Cairns,~I.~H.; McKenzie,~D.~R.
	Deployment Opportunities for Space Photovoltaics and the Prospects for
	Perovskite Solar Cells. \emph{Adv. Mater. Technol.} \textbf{2022}, \emph{7},
	2101059\relax
	\mciteBstWouldAddEndPuncttrue
	\mciteSetBstMidEndSepPunct{\mcitedefaultmidpunct}
	{\mcitedefaultendpunct}{\mcitedefaultseppunct}\relax
	\EndOfBibitem
	\bibitem[Pérez-del Rey \latin{et~al.}(2020)Pérez-del Rey, Dreessen,
	Igual-Muñoz, van~den Hengel, Gélvez-Rueda, Savenije, Grozema, Zimmermann,
	and Bolink]{spaceperovskite}
	Pérez-del Rey,~D.; Dreessen,~C.; Igual-Muñoz,~A.~M.; van~den Hengel,~L.;
	Gélvez-Rueda,~M.~C.; Savenije,~T.~J.; Grozema,~F.~C.; Zimmermann,~C.;
	Bolink,~H.~J. Perovskite Solar Cells: Stable under Space Conditions.
	\emph{Sol. RRL} \textbf{2020}, \emph{4}, 2000447\relax
	\mciteBstWouldAddEndPuncttrue
	\mciteSetBstMidEndSepPunct{\mcitedefaultmidpunct}
	{\mcitedefaultendpunct}{\mcitedefaultseppunct}\relax
	\EndOfBibitem
	\bibitem[Park and Zhu(2020)Park, and Zhu]{perovskitnapelem}
	Park,~N.-G.; Zhu,~K. Scalable fabrication and coating methods for perovskite
	solar cells and solar modules. \emph{Nat. Rev. Mater.} \textbf{2020},
	\emph{5}, 333--350\relax
	\mciteBstWouldAddEndPuncttrue
	\mciteSetBstMidEndSepPunct{\mcitedefaultmidpunct}
	{\mcitedefaultendpunct}{\mcitedefaultseppunct}\relax
	\EndOfBibitem
	\bibitem[Yuan \latin{et~al.}(2015)Yuan, Chae, Shao, Wang, Xiao, Centrone, and
	Huang]{Perovskite_IR_study}
	Yuan,~Y.; Chae,~J.; Shao,~Y.; Wang,~Q.; Xiao,~Z.; Centrone,~A.; Huang,~J.
	{Photovoltaic Switching Mechanism in Lateral Structure Hybrid Perovskite
		Solar Cells}. \emph{Adv. Energy Mater.} \textbf{2015}, \emph{5}\relax
	\mciteBstWouldAddEndPuncttrue
	\mciteSetBstMidEndSepPunct{\mcitedefaultmidpunct}
	{\mcitedefaultendpunct}{\mcitedefaultseppunct}\relax
	\EndOfBibitem
	\bibitem[Bernard \latin{et~al.}(2018)Bernard, Wasylishen, Ratcliffe, Terskikh,
	Wu, Buriak, and Hauger]{allignemnt2}
	Bernard,~G.~M.; Wasylishen,~R.~E.; Ratcliffe,~C.~I.; Terskikh,~V.; Wu,~Q.;
	Buriak,~J.~M.; Hauger,~T. Methylammonium Cation Dynamics in Methylammonium
	Lead Halide Perovskites: A Solid-State NMR Perspective. \emph{J. Phys. Chem.
		A} \textbf{2018}, \emph{122}, 1560--1573\relax
	\mciteBstWouldAddEndPuncttrue
	\mciteSetBstMidEndSepPunct{\mcitedefaultmidpunct}
	{\mcitedefaultendpunct}{\mcitedefaultseppunct}\relax
	\EndOfBibitem
	\bibitem[Tian \latin{et~al.}(2020)Tian, Xue, Yao, Li, Brabec, and
	Yip]{InorganicPerovskiteReview}
	Tian,~J.; Xue,~Q.; Yao,~Q.; Li,~N.; Brabec,~C.~J.; Yip,~H.-L. Inorganic Halide
	Perovskite Solar Cells: Progress and Challenges. \emph{Adv. Energy Mater.}
	\textbf{2020}, \emph{10}\relax
	\mciteBstWouldAddEndPuncttrue
	\mciteSetBstMidEndSepPunct{\mcitedefaultmidpunct}
	{\mcitedefaultendpunct}{\mcitedefaultseppunct}\relax
	\EndOfBibitem
	\bibitem[Hu \latin{et~al.}(2017)Hu, Aygüler, Petrus, Bein, and
	Docampo]{CsPbBr3_mixing_Cs_stable}
	Hu,~Y.; Aygüler,~M.~F.; Petrus,~M.~L.; Bein,~T.; Docampo,~P. {Impact of
		Rubidium and Cesium Cations on the Moisture Stability of Multiple-Cation
		Mixed-Halide Perovskites}. \emph{ACS Energy Lett.} \textbf{2017}, \emph{2},
	2212--2218\relax
	\mciteBstWouldAddEndPuncttrue
	\mciteSetBstMidEndSepPunct{\mcitedefaultmidpunct}
	{\mcitedefaultendpunct}{\mcitedefaultseppunct}\relax
	\EndOfBibitem
	\bibitem[Clinckemalie \latin{et~al.}(2021)Clinckemalie, Valli, Roeffaers,
	Hofkens, Pradhan, and Debroye]{CsPbBr3_scintillator}
	Clinckemalie,~L.; Valli,~D.; Roeffaers,~M. B.~J.; Hofkens,~J.; Pradhan,~B.;
	Debroye,~E. {Challenges and Opportunities for CsPbBr$_3$ Perovskites in Low-
		and High-Energy Radiation Detection}. \emph{ACS Energy Lett.} \textbf{2021},
	\emph{6}, 1290--1314\relax
	\mciteBstWouldAddEndPuncttrue
	\mciteSetBstMidEndSepPunct{\mcitedefaultmidpunct}
	{\mcitedefaultendpunct}{\mcitedefaultseppunct}\relax
	\EndOfBibitem
	\bibitem[Zhou and Zhao(2019)Zhou, and Zhao]{CsPbBr3_stable}
	Zhou,~Y.; Zhao,~Y. Chemical stability and instability of inorganic halide
	perovskites. \emph{Energy Environ. Sci.} \textbf{2019}, \emph{12},
	1495--1511\relax
	\mciteBstWouldAddEndPuncttrue
	\mciteSetBstMidEndSepPunct{\mcitedefaultmidpunct}
	{\mcitedefaultendpunct}{\mcitedefaultseppunct}\relax
	\EndOfBibitem
	\bibitem[Haruta \latin{et~al.}(2020)Haruta, Ikenoue, Miyake, and
	Hirato]{CsPbBr3_napelem}
	Haruta,~Y.; Ikenoue,~T.; Miyake,~M.; Hirato,~T. One-Step Coating of
	Full-Coverage CsPbBr$_3$ Thin Films via Mist Deposition for All-Inorganic
	Perovskite Solar Cells. \emph{ACS Appl. Energy Mater.} \textbf{2020},
	\emph{3}, 11523--11528\relax
	\mciteBstWouldAddEndPuncttrue
	\mciteSetBstMidEndSepPunct{\mcitedefaultmidpunct}
	{\mcitedefaultendpunct}{\mcitedefaultseppunct}\relax
	\EndOfBibitem
	\bibitem[Ding \latin{et~al.}(2017)Ding, Du, Zuo, Zhao, Cui, and
	Zhan]{CsPbBr3_detector_PL_Abs}
	Ding,~J.; Du,~S.; Zuo,~Z.; Zhao,~Y.; Cui,~H.; Zhan,~X. High Detectivity and
	Rapid Response in Perovskite CsPbBr$_3$ Single-Crystal Photodetector.
	\emph{J. Phys. Chem. C} \textbf{2017}, \emph{121}, 4917--4923\relax
	\mciteBstWouldAddEndPuncttrue
	\mciteSetBstMidEndSepPunct{\mcitedefaultmidpunct}
	{\mcitedefaultendpunct}{\mcitedefaultseppunct}\relax
	\EndOfBibitem
	\bibitem[Fu \latin{et~al.}(2016)Fu, Zhu, Stoumpos, Ding, Wang, Kanatzidis, Zhu,
	and Jin]{CsPbBr3_laser}
	Fu,~Y.; Zhu,~H.; Stoumpos,~C.~C.; Ding,~Q.; Wang,~J.; Kanatzidis,~M.~G.;
	Zhu,~X.; Jin,~S. Broad Wavelength Tunable Robust Lasing from Single-Crystal
	Nanowires of Cesium Lead Halide Perovskites (CsPbX3, X = Cl, Br, I).
	\emph{ACS Nano} \textbf{2016}, \emph{10}, 7963--7972\relax
	\mciteBstWouldAddEndPuncttrue
	\mciteSetBstMidEndSepPunct{\mcitedefaultmidpunct}
	{\mcitedefaultendpunct}{\mcitedefaultseppunct}\relax
	\EndOfBibitem
	\bibitem[Eaton \latin{et~al.}(2016)Eaton, Lai, Gibson, Wong, Dou, Ma, Wang,
	Leone, and Yang]{CsPbBr3_laser2}
	Eaton,~S.~W.; Lai,~M.; Gibson,~N.~A.; Wong,~A.~B.; Dou,~L.; Ma,~J.;
	Wang,~L.-W.; Leone,~S.~R.; Yang,~P. Lasing in robust cesium lead halide
	perovskite nanowires. \emph{Proc. Natl. Acad. Sci.} \textbf{2016},
	\emph{113}, 1993--1998\relax
	\mciteBstWouldAddEndPuncttrue
	\mciteSetBstMidEndSepPunct{\mcitedefaultmidpunct}
	{\mcitedefaultendpunct}{\mcitedefaultseppunct}\relax
	\EndOfBibitem
	\bibitem[Stoumpos \latin{et~al.}(2013)Stoumpos, Malliakas, Peters, Liu,
	Sebastian, Im, Chasapis, Wibowo, Chung, Freeman, Wessels, and
	Kanatzidis]{CsPbBr3_photoconductivity}
	Stoumpos,~C.~C.; Malliakas,~C.~D.; Peters,~J.~A.; Liu,~Z.; Sebastian,~M.;
	Im,~J.; Chasapis,~T.~C.; Wibowo,~A.~C.; Chung,~D.~Y.; Freeman,~A.~J.;
	Wessels,~B.~W.; Kanatzidis,~M.~G. Crystal Growth of the Perovskite
	Semiconductor CsPbBr$_3$: A New Material for High-Energy Radiation Detection.
	\emph{Cryst. Growth Des.} \textbf{2013}, \emph{13}, 2722--2727\relax
	\mciteBstWouldAddEndPuncttrue
	\mciteSetBstMidEndSepPunct{\mcitedefaultmidpunct}
	{\mcitedefaultendpunct}{\mcitedefaultseppunct}\relax
	\EndOfBibitem
	\bibitem[Hirotsu \latin{et~al.}(1974)Hirotsu, Harada, Iizumi, and
	Gesi]{CsPbBr3_phases}
	Hirotsu,~S.; Harada,~J.; Iizumi,~M.; Gesi,~K. Structural Phase Transitions in
	CsPbBr$_3$. \emph{J. Phys. Soc. Jpn.} \textbf{1974}, \emph{37},
	1393--1398\relax
	\mciteBstWouldAddEndPuncttrue
	\mciteSetBstMidEndSepPunct{\mcitedefaultmidpunct}
	{\mcitedefaultendpunct}{\mcitedefaultseppunct}\relax
	\EndOfBibitem
	\bibitem[Gabelloni \latin{et~al.}(2017)Gabelloni, Biccari, Andreotti, Balestri,
	Checcucci, Milanesi, Calisi, Caporali, and Vinattieri]{CsPbBr3_tobbcsucs1}
	Gabelloni,~F.; Biccari,~F.; Andreotti,~G.; Balestri,~D.; Checcucci,~S.;
	Milanesi,~A.; Calisi,~N.; Caporali,~S.; Vinattieri,~A. Recombination dynamics
	in CsPbBr$_3$ nanocrystals: role of surface states. \emph{Opt. Mater.
		Express} \textbf{2017}, \emph{7}, 4367--4373\relax
	\mciteBstWouldAddEndPuncttrue
	\mciteSetBstMidEndSepPunct{\mcitedefaultmidpunct}
	{\mcitedefaultendpunct}{\mcitedefaultseppunct}\relax
	\EndOfBibitem
	\bibitem[Yuan \latin{et~al.}(2020)Yuan, Chen, Yang, Shen, Liu, and
	Cao]{CsPbBr3_Tfuggo_PL_ABS}
	Yuan,~Y.; Chen,~M.; Yang,~S.; Shen,~X.; Liu,~Y.; Cao,~D. Exciton recombination
	mechanisms in solution grown single crystalline CsPbBr$_3$ perovskite.
	\emph{J. Lumin.} \textbf{2020}, \emph{226}, 117471\relax
	\mciteBstWouldAddEndPuncttrue
	\mciteSetBstMidEndSepPunct{\mcitedefaultmidpunct}
	{\mcitedefaultendpunct}{\mcitedefaultseppunct}\relax
	\EndOfBibitem
	\bibitem[Lao \latin{et~al.}(2018)Lao, Yang, Su, Wang, Ye, Wang, Yao, and
	Xu]{CsPbBr3_ketcsucs2}
	Lao,~X.; Yang,~Z.; Su,~Z.; Wang,~Z.; Ye,~H.; Wang,~M.; Yao,~X.; Xu,~S.
	Luminescence and thermal behaviors of free and trapped excitons in cesium
	lead halide perovskite nanosheets. \emph{Nanoscale} \textbf{2018}, \emph{10},
	9949--9956\relax
	\mciteBstWouldAddEndPuncttrue
	\mciteSetBstMidEndSepPunct{\mcitedefaultmidpunct}
	{\mcitedefaultendpunct}{\mcitedefaultseppunct}\relax
	\EndOfBibitem
	\bibitem[Liao \latin{et~al.}(2021)Liao, Shan, and Li]{CsPbBr3_TRPL1}
	Liao,~M.; Shan,~B.; Li,~M. Role of Trap States in Excitation
	Wavelength-Dependent Photoluminescence of Strongly Quantum-Confined
	All-Inorganic CsPbBr$_3$ Perovskites with Varying Dimensionalities. \emph{J.
		Phys. Chem. C} \textbf{2021}, \emph{125}, 21062--21069\relax
	\mciteBstWouldAddEndPuncttrue
	\mciteSetBstMidEndSepPunct{\mcitedefaultmidpunct}
	{\mcitedefaultendpunct}{\mcitedefaultseppunct}\relax
	\EndOfBibitem
	\bibitem[Akkerman \latin{et~al.}(2016)Akkerman, Motti, Srimath~Kandada,
	Mosconi, D’Innocenzo, Bertoni, Marras, Kamino, Miranda, De~Angelis,
	Petrozza, Prato, and Manna]{CsPbBr3_TRPL3}
	Akkerman,~Q.~A.; Motti,~S.~G.; Srimath~Kandada,~A.~R.; Mosconi,~E.;
	D’Innocenzo,~V.; Bertoni,~G.; Marras,~S.; Kamino,~B.~A.; Miranda,~L.;
	De~Angelis,~F.; Petrozza,~A.; Prato,~M.; Manna,~L. Solution Synthesis
	Approach to Colloidal Cesium Lead Halide Perovskite Nanoplatelets with
	Monolayer-Level Thickness Control. \emph{J. Am. Chem. Soc.} \textbf{2016},
	\emph{138}, 1010--1016\relax
	\mciteBstWouldAddEndPuncttrue
	\mciteSetBstMidEndSepPunct{\mcitedefaultmidpunct}
	{\mcitedefaultendpunct}{\mcitedefaultseppunct}\relax
	\EndOfBibitem
	\bibitem[Liang \latin{et~al.}(2016)Liang, Zhao, Xu, Qiao, Song, Gao, and
	Xu]{CsPbBr3_TRPL4}
	Liang,~Z.; Zhao,~S.; Xu,~Z.; Qiao,~B.; Song,~P.; Gao,~D.; Xu,~X.
	Shape-Controlled Synthesis of All-Inorganic CsPbBr$_3$ Perovskite
	Nanocrystals with Bright Blue Emission. \emph{ACS Appl. Mater. Interfaces}
	\textbf{2016}, \emph{8}, 28824--28830\relax
	\mciteBstWouldAddEndPuncttrue
	\mciteSetBstMidEndSepPunct{\mcitedefaultmidpunct}
	{\mcitedefaultendpunct}{\mcitedefaultseppunct}\relax
	\EndOfBibitem
	\bibitem[Zhang \latin{et~al.}(2019)Zhang, Guo, Yang, Bose, Liu, Yin, Han, Bakr,
	Mohammed, and Malko]{CsPbBr3_TRPL2}
	Zhang,~Y.; Guo,~T.; Yang,~H.; Bose,~R.; Liu,~L.; Yin,~J.; Han,~Y.; Bakr,~O.~M.;
	Mohammed,~O.~F.; Malko,~A.~V. Emergence of multiple fluorophores in
	individual cesium lead bromide nanocrystals. \emph{Nat. Commun.}
	\textbf{2019}, \emph{10}, 2930\relax
	\mciteBstWouldAddEndPuncttrue
	\mciteSetBstMidEndSepPunct{\mcitedefaultmidpunct}
	{\mcitedefaultendpunct}{\mcitedefaultseppunct}\relax
	\EndOfBibitem
	\bibitem[Zhang \latin{et~al.}(2020)Zhang, Pang, Xing, and Chen]{CsPbBr3_TRPL5}
	Zhang,~X.; Pang,~G.; Xing,~G.; Chen,~R. Temperature dependent optical
	characteristics of all-inorganic CsPbBr$_3$ nanocrystals film. \emph{Mater.
		Today Phys.} \textbf{2020}, \emph{15}, 100259\relax
	\mciteBstWouldAddEndPuncttrue
	\mciteSetBstMidEndSepPunct{\mcitedefaultmidpunct}
	{\mcitedefaultendpunct}{\mcitedefaultseppunct}\relax
	\EndOfBibitem
	\bibitem[Lobo \latin{et~al.}(2022)Lobo, Kawane, Matt, Osvet, Shrestha, Ievgen,
	Brabec, Kanak, Fochuk, and Kato]{CsPbBr3_muPCD}
	Lobo,~N.; Kawane,~T.; Matt,~G.~J.; Osvet,~A.; Shrestha,~S.; Ievgen,~L.;
	Brabec,~C.~J.; Kanak,~A.; Fochuk,~P.; Kato,~M. Trapping effects and
	surface/interface recombination of carrier recombination in single- or
	poly-crystalline metal halide perovskites. \emph{Jpn. J. Appl. Phys.}
	\textbf{2022}, \emph{61}, 125503\relax
	\mciteBstWouldAddEndPuncttrue
	\mciteSetBstMidEndSepPunct{\mcitedefaultmidpunct}
	{\mcitedefaultendpunct}{\mcitedefaultseppunct}\relax
	\EndOfBibitem
	\bibitem[G{\'e}lvez-Rueda \latin{et~al.}(2020)G{\'e}lvez-Rueda, Fridriksson,
	Dubey, Jager, van~der Stam, and Grozema]{CsPbBr3_muPCD2}
	G{\'e}lvez-Rueda,~M.~C.; Fridriksson,~M.~B.; Dubey,~R.~K.; Jager,~W.~F.;
	van~der Stam,~W.; Grozema,~F.~C. Overcoming the exciton binding energy in
	two-dimensional perovskite nanoplatelets by attachment of conjugated organic
	chromophores. \emph{Nat. Commun.} \textbf{2020}, \emph{11}, 1901\relax
	\mciteBstWouldAddEndPuncttrue
	\mciteSetBstMidEndSepPunct{\mcitedefaultmidpunct}
	{\mcitedefaultendpunct}{\mcitedefaultseppunct}\relax
	\EndOfBibitem
	\bibitem[{Kunst} and {Beck}(1986){Kunst}, and {Beck}]{kunst1}
	{Kunst},~M.; {Beck},~G. {The study of charge carrier kinetics in semiconductors
		by microwave conductivity measurements}. \emph{J. Appl. Phys.} \textbf{1986},
	\emph{60}, 3558--3566\relax
	\mciteBstWouldAddEndPuncttrue
	\mciteSetBstMidEndSepPunct{\mcitedefaultmidpunct}
	{\mcitedefaultendpunct}{\mcitedefaultseppunct}\relax
	\EndOfBibitem
	\bibitem[Kunst and Beck(1988)Kunst, and Beck]{kunst2}
	Kunst,~M.; Beck,~G. The study of charge carrier kinetics in semiconductors by
	microwave conductivity measurements. $\text{II}$. \emph{J. Appl. Phys.}
	\textbf{1988}, \emph{63}, 1093--1098\relax
	\mciteBstWouldAddEndPuncttrue
	\mciteSetBstMidEndSepPunct{\mcitedefaultmidpunct}
	{\mcitedefaultendpunct}{\mcitedefaultseppunct}\relax
	\EndOfBibitem
	\bibitem[Bojtor \latin{et~al.}(2022)Bojtor, Kollarics, Márkus, Sienkiewicz,
	Kollár, Forró, and Simon]{BojtorACSPhotonics}
	Bojtor,~A.; Kollarics,~S.; Márkus,~B.~G.; Sienkiewicz,~A.; Kollár,~M.;
	Forró,~L.; Simon,~F. Ultralong Charge Carrier Recombination Time in
	Methylammonium Lead Halide Perovskites. \emph{ACS Photonics} \textbf{2022},
	\emph{9}, 3341--3350\relax
	\mciteBstWouldAddEndPuncttrue
	\mciteSetBstMidEndSepPunct{\mcitedefaultmidpunct}
	{\mcitedefaultendpunct}{\mcitedefaultseppunct}\relax
	\EndOfBibitem
	\bibitem[Guo \latin{et~al.}()Guo, Wang, Lu, Wang, Liu, Wang, Dong, Zhou, Zheng,
	Fu, Xie, Wan, Xing, Chen, and Liu]{Bencetol_tandem}
	Guo,~J.; Wang,~B.; Lu,~D.; Wang,~T.; Liu,~T.; Wang,~R.; Dong,~X.; Zhou,~T.;
	Zheng,~N.; Fu,~Q.; Xie,~Z.; Wan,~X.; Xing,~G.; Chen,~Y.; Liu,~Y. Ultralong
	Carrier Lifetime Exceeding 20 µs in Lead Halide Perovskite Film Enable
	Efficient Solar Cells. \emph{Adv. Mater.} \emph{n/a}, 2212126\relax
	\mciteBstWouldAddEndPuncttrue
	\mciteSetBstMidEndSepPunct{\mcitedefaultmidpunct}
	{\mcitedefaultendpunct}{\mcitedefaultseppunct}\relax
	\EndOfBibitem
	\bibitem[{Sinton} \latin{et~al.}(1996){Sinton}, {Cuevas}, and
	{Stuckings}]{Sinton1}
	{Sinton},~R.~A.; {Cuevas},~A.; {Stuckings},~M. Quasi-steady-state
	photoconductance, a new method for solar cell material and device
	characterization. Conference Record of the Twenty Fifth IEEE Photovoltaic
	Specialists Conference - 1996. 1996; pp 457--460\relax
	\mciteBstWouldAddEndPuncttrue
	\mciteSetBstMidEndSepPunct{\mcitedefaultmidpunct}
	{\mcitedefaultendpunct}{\mcitedefaultseppunct}\relax
	\EndOfBibitem
	\bibitem[Goodarzi \latin{et~al.}(2019)Goodarzi, Sinton, and Macdonald]{sinton2}
	Goodarzi,~M.; Sinton,~R.; Macdonald,~D. Quasi-steady-state photoconductance
	bulk lifetime measurements on silicon ingots with deeper photogeneration.
	\emph{AIP Adv.} \textbf{2019}, \emph{9}, 015128\relax
	\mciteBstWouldAddEndPuncttrue
	\mciteSetBstMidEndSepPunct{\mcitedefaultmidpunct}
	{\mcitedefaultendpunct}{\mcitedefaultseppunct}\relax
	\EndOfBibitem
	\bibitem[Gyüre-Garami \latin{et~al.}(2019)Gyüre-Garami, Blum, Sági, Bojtor,
	Kollarics, Csősz, Márkus, Volk, and Simon]{gyregarami2019ultrafast}
	Gyüre-Garami,~B.; Blum,~B.; Sági,~O.; Bojtor,~A.; Kollarics,~S.; Csősz,~G.;
	Márkus,~B.~G.; Volk,~J.; Simon,~F. Ultrafast sensing of photoconductivity
	decay using microwave resonators. \emph{J. Appl. Phys.} \textbf{2019},
	\emph{126}, 235702\relax
	\mciteBstWouldAddEndPuncttrue
	\mciteSetBstMidEndSepPunct{\mcitedefaultmidpunct}
	{\mcitedefaultendpunct}{\mcitedefaultseppunct}\relax
	\EndOfBibitem
	\bibitem[Lauer \latin{et~al.}(2008)Lauer, Laades, Übensee, Metzner, and
	Lawerenz]{muPCD_lauer}
	Lauer,~K.; Laades,~A.; Übensee,~H.; Metzner,~H.; Lawerenz,~A. {Detailed
		analysis of the microwave-detected photoconductance decay in crystalline
		silicon}. \emph{J. Appl. Phys.} \textbf{2008}, \emph{104}\relax
	\mciteBstWouldAddEndPuncttrue
	\mciteSetBstMidEndSepPunct{\mcitedefaultmidpunct}
	{\mcitedefaultendpunct}{\mcitedefaultseppunct}\relax
	\EndOfBibitem
	\bibitem[Bisquert \latin{et~al.}(2009)Bisquert, Fabregat-Santiago, Mora-Seró,
	Garcia-Belmonte, and Giménez]{tau_dn_derivalt_calculation}
	Bisquert,~J.; Fabregat-Santiago,~F.; Mora-Seró,~I.; Garcia-Belmonte,~G.;
	Giménez,~S. Electron Lifetime in Dye-Sensitized Solar Cells: Theory and
	Interpretation of Measurements. \emph{J. Phys. Chem. C} \textbf{2009},
	\emph{113}, 17278--17290\relax
	\mciteBstWouldAddEndPuncttrue
	\mciteSetBstMidEndSepPunct{\mcitedefaultmidpunct}
	{\mcitedefaultendpunct}{\mcitedefaultseppunct}\relax
	\EndOfBibitem
	\bibitem[Wilson \latin{et~al.}(2011)Wilson, Savtchouk, Lagowski, Kis-Szabo,
	Korsos, Toth, Kopecek, and Mihailetchi]{Feriek_QSSPCD_tau_dn}
	Wilson,~M.; Savtchouk,~A.; Lagowski,~J.; Kis-Szabo,~K.; Korsos,~F.; Toth,~A.;
	Kopecek,~R.; Mihailetchi,~V. QSS-$\mu$PCD measurement of lifetime in silicon
	wafers: advantages and new applications. \emph{Energy Procedia}
	\textbf{2011}, \emph{8}, 128--134\relax
	\mciteBstWouldAddEndPuncttrue
	\mciteSetBstMidEndSepPunct{\mcitedefaultmidpunct}
	{\mcitedefaultendpunct}{\mcitedefaultseppunct}\relax
	\EndOfBibitem
	\bibitem[Bowden and Sinton(2007)Bowden, and Sinton]{dn_tau_connection1}
	Bowden,~S.; Sinton,~R.~A. {Determining lifetime in silicon blocks and wafers
		with accurate expressions for carrier density}. \emph{J. Appl. Phys.}
	\textbf{2007}, \emph{102}\relax
	\mciteBstWouldAddEndPuncttrue
	\mciteSetBstMidEndSepPunct{\mcitedefaultmidpunct}
	{\mcitedefaultendpunct}{\mcitedefaultseppunct}\relax
	\EndOfBibitem
	\bibitem[Shockley and Read(1952)Shockley, and Read]{SRH_Auger_rad}
	Shockley,~W.; Read,~W.~T. Statistics of the Recombinations of Holes and
	Electrons. \emph{Phys. Rev.} \textbf{1952}, \emph{87}, 835--842\relax
	\mciteBstWouldAddEndPuncttrue
	\mciteSetBstMidEndSepPunct{\mcitedefaultmidpunct}
	{\mcitedefaultendpunct}{\mcitedefaultseppunct}\relax
	\EndOfBibitem
	\bibitem[Hall(1959)]{Hall_alapmu}
	Hall,~R. Recombination processes in semiconductors. \emph{Proc. Inst. Electr.
		Eng., Part B} \textbf{1959}, \emph{106}, 923--931(8)\relax
	\mciteBstWouldAddEndPuncttrue
	\mciteSetBstMidEndSepPunct{\mcitedefaultmidpunct}
	{\mcitedefaultendpunct}{\mcitedefaultseppunct}\relax
	\EndOfBibitem
	\bibitem[Trimpl \latin{et~al.}(2020)Trimpl, Wright, Schutt, Buizza, Wang,
	Johnston, Snaith, Mueller-Buschbaum, and Herz]{Organic_Halide_trapping}
	Trimpl,~M.~J.; Wright,~A.~D.; Schutt,~K.; Buizza,~L. R.~V.; Wang,~Z.;
	Johnston,~M.~B.; Snaith,~H.~J.; Mueller-Buschbaum,~P.; Herz,~L.~M.
	{Charge-Carrier Trapping and Radiative Recombination in Metal Halide
		Perovskite Semiconductors}. \emph{Adv. Funct. Mater.} \textbf{2020},
	\emph{30}\relax
	\mciteBstWouldAddEndPuncttrue
	\mciteSetBstMidEndSepPunct{\mcitedefaultmidpunct}
	{\mcitedefaultendpunct}{\mcitedefaultseppunct}\relax
	\EndOfBibitem
	\bibitem[Zhang \latin{et~al.}(2018)Zhang, Zheng, Fu, Guo, Zhang, Chen, Chen,
	Wang, Luo, and Tian]{cspbbr3_trappak_meres}
	Zhang,~M.; Zheng,~Z.; Fu,~Q.; Guo,~P.; Zhang,~S.; Chen,~C.; Chen,~H.; Wang,~M.;
	Luo,~W.; Tian,~Y. Determination of Defect Levels in Melt-Grown All-Inorganic
	Perovskite CsPbBr$_3$ Crystals by Thermally Stimulated Current Spectra.
	\emph{J. Phys. Chem. C} \textbf{2018}, \emph{122}, 10309--10315\relax
	\mciteBstWouldAddEndPuncttrue
	\mciteSetBstMidEndSepPunct{\mcitedefaultmidpunct}
	{\mcitedefaultendpunct}{\mcitedefaultseppunct}\relax
	\EndOfBibitem
	\bibitem[Kang and Wang(2017)Kang, and Wang]{cspbbr3_trapping_elmelet}
	Kang,~J.; Wang,~L.-W. High Defect Tolerance in Lead Halide Perovskite
	CsPbBr$_3$. \emph{J. Phys. Chem. Lett.} \textbf{2017}, \emph{8},
	489--493\relax
	\mciteBstWouldAddEndPuncttrue
	\mciteSetBstMidEndSepPunct{\mcitedefaultmidpunct}
	{\mcitedefaultendpunct}{\mcitedefaultseppunct}\relax
	\EndOfBibitem
	\bibitem[Svirskas \latin{et~al.}(2020)Svirskas, Balčiūnas, Šimėnas,
	Usevičius, Kinka, Velička, Kubicki, Castillo, Karabanov, Shvartsman,
	de~Rosário~Soares, Šablinskas, Salak, Lupascu, and
	Banys]{CsPbBr3_effectat220K}
	Svirskas,~v.; Balčiūnas,~S.; Šimėnas,~M.; Usevičius,~G.; Kinka,~M.;
	Velička,~M.; Kubicki,~D.; Castillo,~M.~E.; Karabanov,~A.; Shvartsman,~V.~V.;
	de~Rosário~Soares,~M.; Šablinskas,~V.; Salak,~A.~N.; Lupascu,~D.~C.;
	Banys,~J. Phase transitions{,} screening and dielectric response of
	CsPbBr$_3$. \emph{J. Mater. Chem. A} \textbf{2020}, \emph{8},
	14015--14022\relax
	\mciteBstWouldAddEndPuncttrue
	\mciteSetBstMidEndSepPunct{\mcitedefaultmidpunct}
	{\mcitedefaultendpunct}{\mcitedefaultseppunct}\relax
	\EndOfBibitem
	\bibitem[Dirin \latin{et~al.}(2016)Dirin, Cherniukh, Yakunin, Shynkarenko, and
	Kovalenko]{CsPbBr3_sampleprepare3}
	Dirin,~D.~N.; Cherniukh,~I.; Yakunin,~S.; Shynkarenko,~Y.; Kovalenko,~M.~V.
	Solution-Grown CsPbBr$_3$ Perovskite Single Crystals for Photon Detection.
	\emph{Chem. Mater.} \textbf{2016}, \emph{28}, 8470--8474\relax
	\mciteBstWouldAddEndPuncttrue
	\mciteSetBstMidEndSepPunct{\mcitedefaultmidpunct}
	{\mcitedefaultendpunct}{\mcitedefaultseppunct}\relax
	\EndOfBibitem
	\bibitem[Saidaminov \latin{et~al.}(2015)Saidaminov, Abdelhady, Maculan, and
	Bakr]{CsPbBr3_sampleprepare1}
	Saidaminov,~M.~I.; Abdelhady,~A.~L.; Maculan,~G.; Bakr,~O.~M. Retrograde
	solubility of formamidinium and methylammonium lead halide perovskites
	enabling rapid single crystal growth. \emph{Chem. Commun.} \textbf{2015},
	\emph{51}, 17658--17661\relax
	\mciteBstWouldAddEndPuncttrue
	\mciteSetBstMidEndSepPunct{\mcitedefaultmidpunct}
	{\mcitedefaultendpunct}{\mcitedefaultseppunct}\relax
	\EndOfBibitem
	\bibitem[Saidaminov \latin{et~al.}(2015)Saidaminov, Abdelhady, Murali,
	Alarousu, Burlakov, Peng, Dursun, Wang, He, Maculan, Goriely, Wu, Mohammed,
	and Bakr]{CsPbBr3_sampleprepare2}
	Saidaminov,~M.~I.; Abdelhady,~A.~L.; Murali,~B.; Alarousu,~E.; Burlakov,~V.~M.;
	Peng,~W.; Dursun,~I.; Wang,~L.; He,~Y.; Maculan,~G.; Goriely,~A.; Wu,~T.;
	Mohammed,~O.~F.; Bakr,~O.~M. High-quality bulk hybrid perovskite single
	crystals within minutes by inverse temperature crystallization. \emph{Nat.
		Commun.} \textbf{2015}, \emph{6}, 7586\relax
	\mciteBstWouldAddEndPuncttrue
	\mciteSetBstMidEndSepPunct{\mcitedefaultmidpunct}
	{\mcitedefaultendpunct}{\mcitedefaultseppunct}\relax
	\EndOfBibitem
	\bibitem[Simons(2001)]{CPWterek}
	Simons,~R.~N. \emph{Coplanar Waveguide Circuits Components \& Systems}, 1st
	ed.; Wiley-IEEE Press, 2001\relax
	\mciteBstWouldAddEndPuncttrue
	\mciteSetBstMidEndSepPunct{\mcitedefaultmidpunct}
	{\mcitedefaultendpunct}{\mcitedefaultseppunct}\relax
	\EndOfBibitem
	\bibitem[Pozar(2011)]{pozar}
	Pozar,~D. \emph{Microwave Engineering - Solutions Manual}; Wiley, 2011; Vol. 4
	ed.\relax
	\mciteBstWouldAddEndPunctfalse
	\mciteSetBstMidEndSepPunct{\mcitedefaultmidpunct}
	{}{\mcitedefaultseppunct}\relax
	\EndOfBibitem
\end{mcitethebibliography}
\providecommand{\latin}[1]{#1}
\makeatletter
\providecommand{\doi}
{\begingroup\let\do\@makeother\dospecials
	\catcode`\{=1 \catcode`\}=2 \doi@aux}
\providecommand{\doi@aux}[1]{\endgroup\texttt{#1}}
\makeatother
\providecommand*\mcitethebibliography{\thebibliography}
\csname @ifundefined\endcsname{endmcitethebibliography}
{\let\endmcitethebibliography\endthebibliography}{}

\end{document}